%% file: main.tex
\documentclass[manuscript,screen, sigconf]{acmart}
\AtBeginDocument{%
  \providecommand\BibTeX{{%
    \normalfont B\kern-0.5em{\scshape i\kern-0.25em b}\kern-0.8em\TeX}}}

\copyrightyear{2024}
\acmYear{2024}
\setcopyright{acmlicensed}\acmConference[SIGCSE Virtual 2024]{Proceedings of the 2024 ACM Virtual Global Computing Education Conference V. 1}{December 5--8, 2024}{Virtual Event, NC, USA}
\acmBooktitle{Proceedings of the 2024 ACM Virtual Global Computing Education Conference V. 1 (SIGCSE Virtual 2024), December 5--8, 2024, Virtual Event, NC, USA}
\acmDOI{10.1145/3649165.3690128}
\acmISBN{979-8-4007-0598-4/24/12}
\acmConference[SIGCSE Virtual'24]{SIGCSE VIRTUAL}{Dec 05--07, 2024}{Virtual}
\acmBooktitle{SIGCSE Virtual'24, Dec 5-7, 2024} 
\acmPrice{}
\acmISBN{}
\usepackage[font=small,skip=0pt]{caption}
\usepackage{tikz}
\usepackage{pgf-pie}
\usepackage{pgfplots}

\usepackage{enumitem}
\usepackage{hyperref}
\begin{document}

\title{FlashHack: Reflections on the Usage of a Micro Hackathon as an Assessment Tool in a Machine Learning Course}

%FlashHack: Reflections on the Usage of Learning by Doing Micro Hackathon as an Assessment Tool in a Machine Learning Course

%Learning by Doing: Reflections on the usage of Hackathon as an Assessment Tool in a Machine Learning Course

%% Of note is the shared affiliation of the first two authors and the
 
%%\author{P D Parthasarathy}\email{p20210042@goa.bits-pilani.ac.in}\orcid{0000-0002-8723-2407}
%%\affiliation{%
%%\institution{BITS Pilani, KK Birla Goa Campus}
%%\city{}\state{Goa}
%%\country{India}
%%\postcode{403726}
%%}

\author{R Indra}\email{indra.ise@bmsce.ac.in}\orcid{0000-0002-1317-9164}
\affiliation{%
  \institution{Department of Information Science and Engineering \\ B.M.S. College of Engineering, Bengaluru, India}
  \city{}\state{}
  \country{}
  \postcode{}
}
\author{P D Parthasarathy}\email{p20210042@goa.bits-pilani.ac.in}\orcid{0000-0002-8723-2407}
\affiliation{%
\institution{Department of Computer Science and Information Systems \\ BITS Pilani, KK Birla Goa Campus, Goa, India}
\city{}\state{}
\country{}
\postcode{}
}
\author{Jatin Ambasana}\email{jatinambasana@gmail.com}\orcid{0009-0008-7986-9265}
\affiliation{%
  \institution{Centre for Educational Technology \\ Indian Institute of Technology Bombay}
  \city{Mumbai}\state{}
  \country{India}
  \postcode{Code}
}
\author{Spruha Satavlekar}\email{spruhasumukh@iitb.ac.in}\orcid{0000-0001-8448-5815}
\affiliation{%
  \institution{Centre for Educational Technology \\ Indian Institute of Technology Bombay}
  \city{Mumbai}\state{}
  \country{India}
  \postcode{Code}
}

\renewcommand{\shortauthors}{}

\begin{abstract}
    \input{abstract}
\end{abstract}



\newenvironment{myquote}%
  {\list{}{\leftmargin=0.1in\rightmargin=0.1in}\item[]}%
  {\endlist}


\maketitle

\input{intro}

\input{relatedWork}

\input{method}

\input{findings}

\input{limitations}

\input{conclusion}

\bibliographystyle{ACM-Reference-Format} 

\bibliography{main.bib}

% \section{Appendices}

% Uncomment if required.

\end{document}

%% file: abstract.tex
Machine learning (ML) course for undergraduates face challenges in assessing student learning and providing practical exposure. Group project-based learning, an increasingly popular form of experiential learning in CS education, encounters certain limitation in participation and non-participation from a few students. Studies also suggest that students find longer programming assignments and project-based assessments distracting and struggle to maintain focus when they coincide with other courses. To tackle these issues, we introduced FlashHack: a monitored, incremental, in-classroom micro Hackathon that combines project-based learning with Hackathon elements. Engaging 229 third year CS undergraduate students in teams of four, FlashHack prompted them to tackle predefined challenges using machine learning techniques within a set timeframe. Assessment criteria emphasized machine learning application, problem-solving, collaboration, and creativity. Our results indicate high student engagement and satisfaction, alongside simplified assessment processes for instructors. This \emph{experience report} outlines the Hackathon design and implementation, highlights successes and areas for improvement making it feasible for replication by interested computing educators. 

%% file: intro.tex
% !TeX root = main.tex
\section{Introduction}
\label{sec:intro}

Hackathons have emerged as popular events in the tech community \cite{6809711,iraniHackathonsMakingEntrepreneurial2015}, offering participants the opportunity to collaborate, innovate, and develop solutions to real-world challenges within a condensed timeframe \cite{saraviSystemsEngineeringHackathon2018a}. These events are characterized by their fast-paced, hands-on nature, where individuals or teams work intensively to prototype ideas, experiment with new technologies, and demonstrate their creativity and problem-solving skills.

In recent years, hackathons have expanded beyond their traditional domains and have been increasingly utilized as educational tools in various academic settings \cite{10.1145/3197091.3197138}. Recognizing the value of experiential learning and the need to provide students with practical, real-world experiences, educators have embraced hackathons to enhance student engagement, foster interdisciplinary collaboration, and promote innovation in the classroom.

However, on one hand, technical hackathons in curriculum framework may lack structure, guidance, and support, making it difficult for students to navigate the complexities of the event and achieve meaningful learning outcomes. Hackathons, while fostering creativity and rapid prototyping, possess few drawbacks that limit their effectiveness, particularly in the realm of computer science education. One significant drawback is emphasizing short-term, intense collaboration, which may not adequately reflect the skills needed for sustained project development and collaboration in real-world scenarios \cite{uffreduzzi2016hackathon}. Moreover, hackathons often prioritize quantity over quality, leading to hastily developed solutions in short time which may lack robustness and scalability.

% On para on experiential learning in CS currently and its limitations in the Indian context and how a hackathon facilitates experiential learning.

On the other hand, there are Marathon-style hackathons \cite{rice2015hackathon} running over long durations. However, \cite{8747149} indicates that the majority of students perceive that programming assignments and project-based assessments extending beyond a week tend to coincide with other subjects, leading to distraction and reduced focus. While these assessments undeniably enhance skills like time management, they may not always be conducive to courses introducing entirely new material. In order to provide students with experiential, project-based learning without imposing lengthy timelines, we utilize a customized version of a hackathon.

We propose a \emph{FlashHack}: a monitored, incremental,
in-classroom micro hackathon to address these challenges and maximize the educational potential of hackathon. The FlashHack combines traditional hackathons' collaborative, hands-on nature with structured guidance and support, creating an environment that promotes learning, experimentation, and innovation while ensuring participants receive the necessary guidance and support to succeed. We also discuss how this approach is helpful to the educator in quick and efficient assessments. 

In this paper, we introduce the concept of the FlashHack and discuss its potential benefits for student learning and engagement. We outline the key components of this hackathon, including the role of instructors, the structure of the event, and the assessment framework. We present our findings after implementing it in a machine learning course for undergraduate students, highlighting student experiences and learning outcomes. We also comment on what worked well, what could have been better, and provide recommendations to practitioners for future replications. 

This paper is organized as follows: Section \ref{sec:relatedwork} briefly describes the existing literature and how our work contributes to computing education. Section \ref{sec:method} details the course, participants, implementation of the hackathon, and the various assessment instruments used. Our findings are presented and discussed in Section \ref{sec:findings}, highlighting successes and areas of improvement. The paper is concluded in Section \ref{sec:conclusion}.  

%with . Section \ref{sec:recommendations} offers recommendations to computing educators interested in using hackathons as assessments. Section \ref{sec:limitations} describes the limitations of our work and the paper is concluded in Section \ref{sec:conclusion}.

%% file: relatedWork.tex
% !TeX root = main.tex
\section{Related Work}
\label{sec:relatedwork}

\subsection{Hackathons}
Hackathons, commonly known as coding marathons or innovation competitions, are intensive events where individuals or teams collaborate to develop creative solutions to specific challenges within a limited timeframe, typically ranging from a few hours to a few days \cite{6809711}. These events originated in the tech industry but have since gained popularity across various fields, including computer science, engineering, business, and social innovation \cite{9463211}. Hackathons provide participants a platform to showcase their skills, experiment with new technologies, network with peers and industry professionals, and tackle real-world problems in a fast-paced and supportive environment \cite{10.1145/3415216,7985658,8399740}. Studies suggest that the primary motivation driving individuals to engage in hackathons is the chance to acquire new knowledge, with learning taking place through collaborative and situated experiences \cite{10.1007/978-3-662-44426-9_23,inproceedingsCrowd}. This motivation has led to the widespread adoption of hackathons not only within enterprises and corporations but also in the realms of computer science and software engineering education. We elaborate on this further in the subsequent section.

\subsection{Hackathons in Computing Education}
Aiming to create new and engaging learning experiences, computing educators have adopted hackathons in various formal and informal ways \cite{10.1145/2839509.2844590,10.1145/3422392.3422479,engineeringHackathons}. Gama et al. \cite{10.1145/3197091.3197138} introduce hackathons in the formal learning process in an undergraduate Internet of Things (IoT) course and find the practice highly effective. On the other hand, Nandi et al. \cite{10.1145/2839509.2844590} use hackathons in an informal setup as an annual university event and observe that despite the competitive nature, a significant amount of peer learning occurred where students taught each other how to solve specific challenges and learn new skills. Porras et al \cite{10.1145/3194779.3194783} conducted a systematic literature review on utilizing hackathons in software engineering education spanning a decade of events. They propose a taxonomy drawn from their observations to assist practitioners in determining the most suitable intensive event approach based on their industry and educational requirements. Hackathons have also effectively promoted computer programming skills among junior computing students \cite{9334089}. Bolatzhan et al. \cite{8747149} use hackathons in a software engineering course as a project-based teaching tool and find the approach effective and welcoming by students. 

What distinguishes our approach from existing literature on hackathons in computing education is, to the best of our knowledge, the novelty of this being the first experience report documenting the utilization of hackathons in a machine learning course, particularly as a formal assessment instrument.

%% file: method.tex
% !TeX root = main.tex
\section{About the Course}
\label{sec:method}
This section outlines the Machine Learning (ML) course, participant particulars, and details of the Hackathon's implementation and the assessment instruments used. 

\subsection{The Course}
The Machine Learning (ML) course is a mandatory, four-credit course for our university's third-year computer science engineering students. The topics include basic Supervised and Unsupervised ML algorithms such as Classification, Regression, Clustering and Dimensionality reduction. The course consists of 3 hours of lectures and 2 hours of practical sessions every week. The following are some of the course's key Learning Outcomes (LOs)\cite{bloomsTaxonomy}: \\
The learners should be able to -
\begin{itemize}
%     \item LO1: To \textbf{Understand} and acquire knowledge on basic concepts of Machine Learning
% techniques such as supervised and unsupervised learning.
%     \item LO2: To \textbf{Apply} the concepts of Classification, Regression, Clustering and
% Dimensionality reduction algorithms to a given problem.
    \item LO1: Explain and summarize an ML algorithm.
\item LO2: Demonstrate the implementation of ML algorithms
using EdTech tools.
    \item LO3: Determine ML techniques suitable for a given
problem.
\item LO4: Apply teamwork and communication skills to solve an applied ML problem.
\item LO5: Evaluate multiple model results to optimize solutions.
\end{itemize}
The instructional approach incorporated live coding and data analysis and visualization during the lecture sessions. Students were assigned to explore Python libraries for programming machine learning models and given sample datasets to practice their skills in model training and testing after each session. The course followed a comprehensive assessment framework consisting of both Continuous Internal Assessment (CIA) and a final examination, each contributing 50 marks. The CIA encompassed two components: two tests contributing 25 marks in total and a lab component worth another 25 marks. Notably, the lab test was conducted, allocating 10 marks, while the Hackathon evaluation constituted 15 marks. The course was offered in four sections, led by the primary instructor, the first author of this work, alongside three co-instructors, each handling one section.

\subsection{Participants}
The course had 235 enrollments, six students unable to participate in the Hackathon due to medical reasons. Out of the remaining 229 students, all actively participated in the Hackathon, a mandatory aspect of their CIA. The students were grouped into teams of four individuals, resulting in 58 teams participating in the Hackathon. All participants were Indian nationals aged between 19 and 21. Among them, 171 are male and 58 are female.

\subsection{The Hackathon}
\begin{figure}[h]
\centering
    \hspace*{-5mm}
    \vspace{-4mm}
    \includegraphics[scale=0.75]{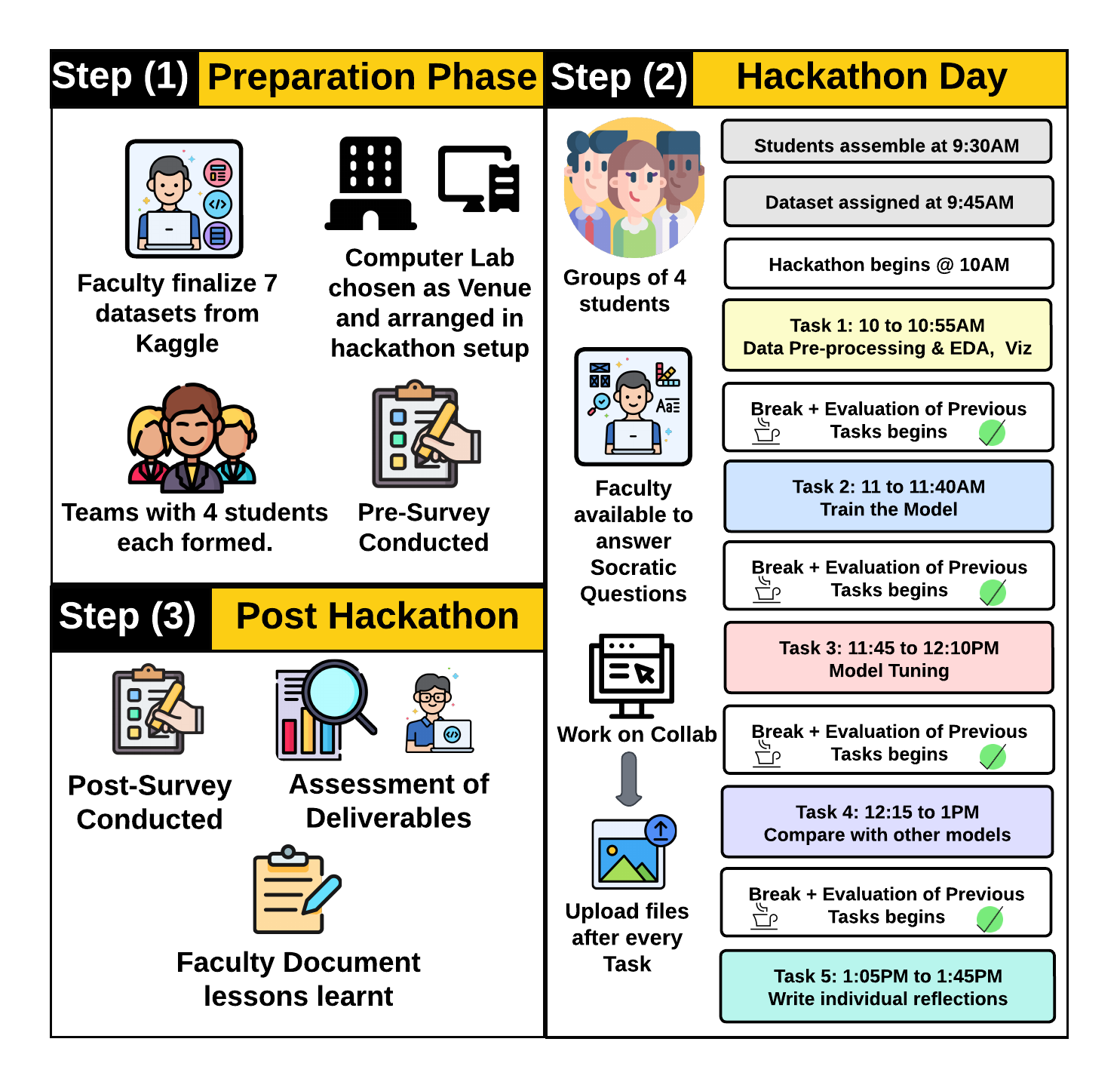}
    \caption{\centering The FlashHack Process with Details of each Phase}
    \label{fig:FlashHackProcess}
    \vspace{-8mm} %5mm vertical space
\end{figure}

As mentioned in the Section \ref{sec:intro}, due to the limitations of the traditional Hackathon, we employ a customized version of the traditional Hackathon as an assessment tool near the conclusion of the semester. This approach involved a supervised, incremental, in-classroom micro Hackathon, which we refer to as \textit{FlashHack}. In this section, we elaborate on all aspects related to FlashHack. The objectives of the FlashHack were explicitly outlined as follows: 
\begin{enumerate}
    \item Foster hands-on experience utilizing machine learning techniques to solve real-world challenges.
    \item Implement theoretical comprehension of ML models.
    \item Enhance problem-solving skills through experiential learning.
    \item Cultivate self-assurance and leadership capabilities among students.
    \item Provide simplified assessment method for ML coursework.
\end{enumerate}
As depicted in Figure \ref{fig:FlashHackProcess}, the FlashHack comprised three main phases: the preparation phase (Pre Hackathon), the FlashHack day, and the Post Hackathon phase. 

\subsubsection{Step 01: Preparation Phase}
The preparation phase aimed to lay the groundwork for the event and build anticipation among students, ensuring they viewed it as an engaging opportunity rather than solely as an assessment. During this phase, the instructors systematically outlined the sequence of activities for the FlashHack day and finalized the datasets to be utilized by teams. Seven datasets of moderate complexity were chosen. FlashHack was scheduled and announced fourteen days in advance. Students were tasked with forming their own teams of four members each and submitting team names and member details via a Google Sheet. Four computer labs were designated as the venue for the FlashHack with 70 capacity each, arranged in a setup resembling a traditional Hackathon to facilitate collaboration within each group.  A pre-survey was administered a week before the event to gauge student interest. The key questions of the survey are provided below for the reader's benefit:

\begin{enumerate}
    \item \label{PreQ1} Have you participated in any Hackathon previously? \textit{[Yes/No]}
    \item \label{PreQ2} What is your opinion on the upcoming FlashHack? \textit{[I am confident to complete the deliverables / It is my first time, I am nervous / It is my first time, I am excited / I guess it will be challenging yet fun / I don't think this would work well]}
\end{enumerate}

\vspace{1mm}

%%\vspace{1mm}
\subsubsection{Step 02: FlashHack Day}
The FlashHack day commenced at 9:30 AM as students gathered at the venue, organized themselves into groups, and took seats. As a Micro-Hackathon, FlashHack extended over a 4-hour period during which participants were required to accomplish specific tasks within the allotted time frames. Each team was tasked with delivering four outputs after each segment, with each student required to submit a reflective report detailing their contributions as the fifth deliverable.

Throughout the Hackathon, the duration and assessment process were prominently displayed on a large screen for all participants to refer to. The instructor distributed the datasets via Google Classroom to teams at 9:45 AM, formally initiating FlashHack at 10 AM. A total of 58 teams participated, with seven or eight teams being given the same dataset. Utilizing Google Colab for the event, students were provided internet access and authorized to employ Python libraries. Adhering to academic integrity, students were obligated to provide citations for any information sources utilized within their colab text section.

Given the incremental nature of the Micro-Hackathon, participants were presented with 5 time-bound tasks, outlined as follows:

\begin{itemize}
    \item \textbf{Task 1 (10:00 to 10:55 AM):} During this phase, participants focused on data pre-processing, conducting exploratory data analysis, and creating visualizations based on the provided dataset. 
    \item \textbf{Task 2 (11:00 to 11:40 AM):} In this segment, participants were tasked with creating and training a suitable machine learning model based on the processed data.
    \item \textbf{Task 3 (11:45 to 12:10 PM):} Participants dedicated this time to fine-tuning their model and optimizing its parameters to enhance performance.
    \item \textbf{Task 4 (12:15 to 01:00 PM):} During this interval, participants implemented additional machine learning models and compared their results with those of the initial model.
    \item \textbf{Task 5 (01:05 to 01:45 PM):} The final task involved writing individual reflection reports detailing their solution, including their contributions and insights gained throughout the FlashHack event. Reports included details on data cleaning procedures, model fitting, parameter tuning, and model comparison to select the most suitable solution for the provided dataset.
\end{itemize}

Upon completion of each task, students were instructed to upload their Python notebook to the designated secure Google Classroom folder assigned to their team. Assessments commenced simultaneously as students moved on to the next task. Due to the presence of seven datasets and only four instructors available, three additional lab instructors assisted with the evaluation process. Each dataset was evaluated by one instructor. Each deliverable was assessed based on predetermined rubrics, as outlined in Table \ref{tab:rubrics}. The final project submission, along with individual reports of each team, were uploaded to the Google Classroom as a PDF file.

\begin{table}[thbp]
  \centering
  \caption{FlashHack Rubrics\label{tab:rubrics}}
  \begin{tabular}{p{.80\columnwidth}r}
    \toprule
    Evaluation Criteria                                                        & Marks(15) \\\midrule
    1. Data visualization and preprocessing                                       & 4    \\
    2. Identification of appropriate ML model for training                       & 2    \\
    3. Model evaluation and Tuning- Performance metrics \& hyperparameter tuning  & 3   \\
    4. Compare and justify the selection of the most suitable ML model for the dataset, including the identification of an alternative model and its justification               & 3    \\
   5. Individual creativity, collaboration efforts, Reflection report          & 3    \\ \bottomrule
  \end{tabular}
  \vspace{-6mm}
\end{table}
 \vspace{2mm}
The aim was to streamline the assessment process and facilitate prompt evaluation of deliverables. Instructors utilized Socratic questioning techniques to guide teams if they veered off course. For instance, if a dataset necessitated a classification model and a team seemed unsure after a significant portion of the allotted time had passed, the instructor would pose 3-5 Socratic questions such as \textit{"What is the dependent and target variable in the dataset?"} and \textit{"Have you considered the type of data?"}. These questions prompted teams to reassess their approach and make necessary adjustments.

Additionally, students could seek hints or clarification through boolean questions from the instructor when they encountered significant challenges. This approach ensured that teams remained on track and avoided major misconceptions throughout the Hackathon. Students demonstrated enthusiasm despite the probing questions and the instructor's covert evaluation. This timely feedback spurred student engagement and instilled a stronger dedication to the ongoing task.
 
\subsubsection{Step 03: Post Hackathon}
Following the conclusion of FlashHack, a Post-Hackathon phase was initiated to assess the outcomes and gather valuable insights. During this phase, a post-survey was conducted to collect participant feedback regarding their experience. The core part of this survey is presented below: 

\begin{enumerate}
    \item The tasks expected to be completed during the Hackathon were achievable within the allocated time frame. \textit{[Yes/No], Single-Select}.
    \item How was your experience in FlashHack? \textit{Very Good / Moderate / Not Good; Single-select.}
    \item Elaborate your experience in FlashHack. \textit{Short-answer}
    \item Do you think Hackathon gave you some essence of exploring your problem-solving skills? \textit{[Yes/No/Unsure], Single-Select}.
    \item Did your experience in FlashHack help you evaluate your conceptual understanding of the concepts learnt in the course? \textit{[Yes/No/Unsure], Single-Select}.
    \item How helpful was the FlashHack in preparing you to perform better in SEE (final exam)? \textit{5-point Likert-scale}.
    \item In your opinion, was the FlashHack an appropriate assessment for the Machine Learning course?  \textit{[Yes/No/Unsure], Single-Select}.
    \item Write your reflection on your experience in FlashHack. \textit{Short-answer, Optional}
    \item Provide suggestions for improving the FlashHack. \textit{Short-answer, Optional}
\end{enumerate}

All deliverables produced by the teams during the Hackathon were meticulously evaluated on the same day. This ensured timely feedback and allowed for immediate identification of strengths and areas for improvement.

Additionally, instructors documented key lessons learned from organizing and executing the FlashHack. This documentation included insights into the effectiveness of the assessment methodology, and challenges encountered during the event are explained in Section \ref{sec:findings}. These lessons served to inform the continuous improvement of the Hackathon format and ensure its ongoing success as a valuable educational tool.

%% file: findings.tex
% !TeX root = main.tex
\section{Findings and Discussion}
\label{sec:findings}

\subsection{Findings}

% use data from jatin's sheet - basic stats, range vs marks chart curve, box plot for male, female 
% use quotes from qual responses 

In this section, we present the findings from FlashHack. 

\subsubsection{Pre-Survey} 
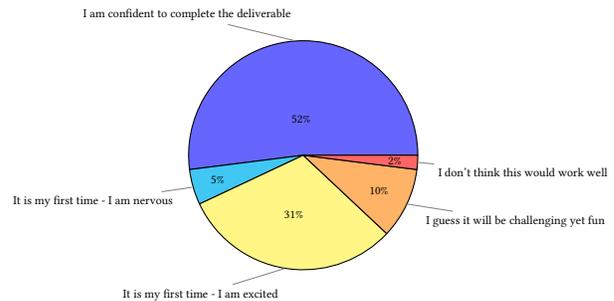
\begin{figure}[h]
    \centering
    \resizebox{0.45\textwidth}{!}{%
    \begin{tikzpicture}
        \pie[text=pin]{52/ I am confident to complete the deliverable, 5/ It is my first time - I
am nervous, 31/It is my first time - I am excited, 10/I guess it will be challenging yet fun,  2/I don’t think this would work well}
    \end{tikzpicture}%
    }
    \caption{Responses to Q2 (Opinion of FlashHack) of the Pre-Survey}
    \label{fig:opinion}
    \vspace{-2mm}
\end{figure}

In response to one of the Question, 58.3\% of respondents reported no prior participation in any Hackathon before FlashHack. Illustrated in Fig. \ref{fig:opinion}, attitudes towards FlashHack varied significantly. A notable majority 52\% expressed confidence in completing the deliverables, while 31\% conveyed excitement, particularly as it was their first experience. Additionally, 10\% anticipated the challenge with enthusiasm, while a smaller fraction 5\% admitted to feeling nervous. Only 2\% expressed skepticism about the event's efficacy.

\subsubsection{Post-Survey}
\begin{figure}[H]
\centering
    %\hspace*{-5mm}
    \vspace{-4mm}
    \includegraphics[scale=0.45]{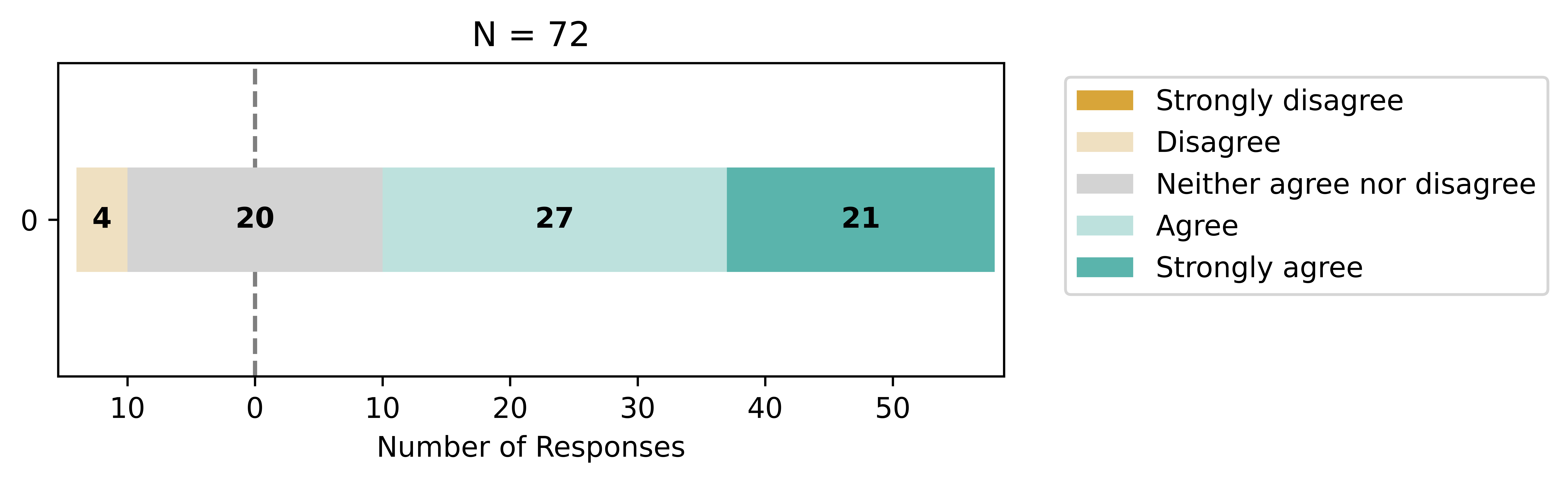}
    \caption{\centering FlashHack helped me prepare and perform better in final exam}
    \label{fig:likertResponse}
    \vspace{-3mm} %5mm vertical space
\end{figure}

Students had to fill the FlashHack Feedback and Survey form after the Hackathon. Albeit instructors mentioned clearly to students that it was not mandatory and will not affect their Hackathon evaluation, it was filled by 72 participants. Out of which, 93.1\% of respondents indicated that the tasks could be accomplished within the allotted time frame. As for their experience with FlashHack, 68.1\% rated it as `Very Good,' while 31.9\% rated it as `Good.' It's worth mentioning that none of the participants expressed a negative opinion. Every participant agreed that the Hackathon provided them with an opportunity to delve into their problem-solving abilities. Moreover, 98\% of respondents expressed that FlashHack aided in assessing their understanding of the course material conceptually, affirming its suitability as an assessment method for the ML course. As depicted in the Likert responses (see \ref{fig:likertResponse}), when queried about whether the Hackathon aided them in preparing and performing better in the final exam, a notable 66.66\% of the respondents (N=72) provided a positive response on the Likert scale and none of them strongly disagreed to the same.

\subsubsection{FlashHack Performance} 
\begin{figure}[h]
\centering
    \vspace{-1mm}
    \includegraphics[scale=0.4]{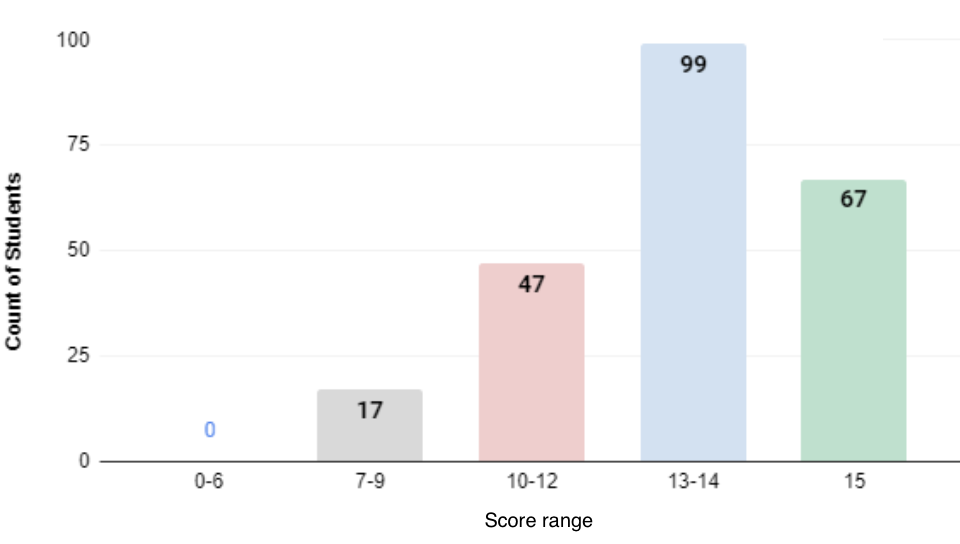}
    \caption{FlashHack Score distribution}
    \label{fig:flashHackScores}
    \vspace{-2mm}
\end{figure}

The FlashHack assessment yielded an average score of 13.1, with a median of 13 and a standard deviation of 1.74. Gender-based analysis revealed no significant difference in performance between male and female students. The distributions of the scores are depicted in Figure \ref{fig:flashHackScores}. Specifically, 67 students achieved the maximum possible score 15, while 99 fell within the 13 to 14 range. Additionally, 47 students scored between 10 and 12, and 17 scored between 7 and 9. Notably, no student scored below 7. Misinterpretation of the data was found to be the reason behind 17 students low performance, with scores ranging from 7 to 9. The data was trained using a regressor model rather than a classifier, leading to incorrect results.

\subsubsection{Qualitative Feedback on FlashHack} We analyzed the responses to short-answer questions from the post-survey using thematic analysis as outlined by Saldaña \cite{saldana_coding_2021}. This involved reading the responses, assigning intial \textsc{Codes}, searching of patterns among the codes to create \textsc{Themes}. The themes were named to reflect their essence clearly and the procedure was repeated twice. Our first short answer question in the Post-survey was to describe the students experience with FlashHack. After performing the two cycles of thematic analysis, we arrived at six themes as explained below. 

\begin{enumerate}
    \item \textbf{Positive Experience:} This theme reflects the overall positive feedback and the educational benefits participants experienced during the Hackathon. Participant P1 reported FlashHack to be \textit{"Smooth and interesting.."} while P18 reported it was \textit{"Helpful and a very good learning experience.."}. P14 reported exceeding his expectations by quoting \textit{"..It was my first time in Hackathon and I performed better than expected and we solved the problem.."}
    \item \textbf{First-time Experience and Skill Development:} This theme captured the experiences of first-time participants and the development of problem-solving skills. Participants enjoyed their first micro Hackathon with most of them reporting that the Hackathon aided them to learn something new. P25 said \textit{"It was my first-time and it [FlasHack] helped to explore my problem solving skills"}. One of the participants P9 also attributed that the FlashHack covered concepts learnt in the class which helped her perform faster in the FlashHack by reporting \textit{"... I was able to complete the Hackathon very fast. It was related to whatever we have learnt in class.."} 
    \item \textbf{Challenges:} Includes the challenges faced by participants and the practical application of their skills and knowledge. Participant P2 states \textit{"It was a challenging task"} while P39 claimed \textit{"Some errors made us think about all the possibilities of getting the error."} Few students, in spite of not able to complete the deliverables, stated learning new things from their peers during the Hackathon. One of the participants stated \textit{"We didn't manage to get output but I did learn stuff!"}. Despite these challenges, the event was seen as an opportunity to build resilience and improve problem-solving skills. Participant stated\textit{".. got chance to build a project from scratch"}.
    \item \textbf{Organization and Time Management:} This theme covers feedback on the Hackathon’s organization, setup, and time constraints faced by participants. While most of them agreed that the organization was done very well, there were mixed responses in the time provided for each deliverable in the FlashHack. One of the students P38 reported \textit{"The Hackathon was conducted in a proper manner, instructions were given clearly therefore the conduction of Hackathon was very good."} 
    Few of them felt the time constraints made the tasks more challenging while others felt it was apt. Participant P50 said \textit{"It was conducted in a proper manner"} while others such as P43 reported \textit{"Lot of time was needed. Might have given more"} and P60 stated \textit{"Time constraints on improving the performance measures"} for model tuning. 
    \item \textbf{Teamwork and Collaboration:} The Hackathon fostered a collaborative environment where participants could work together, share ideas, and solve problems collectively. Responses such as \textit{"Group discussions helped me in solving doubts and increase confidence"} and \textit{"It was a great chance to work with my other mates to come up with a solution"} underscored the importance of teamwork. Participants appreciated the opportunity to work in teams, which not only facilitated better learning but also improved their ability to tackle challenges under pressure. One participant shared, \textit{"I got to work as a team and had explored various things and I have learnt teamwork and how to work under pressure."}Many participants highlighted the collaborative aspects of FlashHack and the importance of team work. Participant P72 said \textit{"It would have been hard for me to do it all by myself. Having friends helped."} while another participant quoted their team effort was exciting by stating \textit{"It was very exciting as the team had to train the model in a certain amount of time."}
    \item \textbf{Learning and Knowledge Gain:} Participants consistently highlighted the educational benefits of the Hackathon, emphasizing their enhanced understanding of machine learning concepts and practical applications. Many noted how the event helped them gain \textit{"powerful insights on Machine Learning and its landscape"} and provided a \textit{"wholesome learning experience in inculcating practical knowledge."} The Hackathon was described as a \textit{"real-time problem-solving experience"} that allowed for \textit{"hands-on learning"} and deepened their understanding, as one participant noted, \textit{"Before Hackathon, my understanding of ML was 60\%. After Hackathon I understood the entire process."}
\end{enumerate}

The last post-survey question asked students for suggestions for improvement in conducting FlashHack. Though we got limited responses to this question, the only theme arising from it was the time duration, where students suggested having a longer duration, facilitating a sense of a true Hackathon rather than a micro Hackathon. 

\subsubsection{Instructor Reflections}
The primary instructor who is the first author led the other section instructors in implementing FlashHack. Reflecting on the Hackathon, it demonstrated a novel way to involve passive students in experiential learning. The assessment strategy and rubrics designed in FlashHack also provided quick and easy to orchestrate means of CIA. The issues of hints and Socratic questions provided to some student groups were discussed among the section heads, post Hackathon.  The reflection noted that providing hints can lead to students excessively relying on the hints from instructor. Potential solution to mitigate this issue was discussed and we present it in the section 4.2.2, \emph{"What could be improved?"}

\subsection{Discussion}
Most students were given a platform to engage in and fully experience the Hackathon actively. Since this was the first time it was being implemented in the department, instructors needed to establish the method. Identifying a suitable dataset was challenging, as it had to allow feasible model training within the specified time frame. To meet this objective, we selected less complex datasets.

\subsubsection{What Worked?} %FlashHack, being done for the first time, is eagerly anticipated for its success.}
\begin{itemize} 
    \item In project-based assignments, there are often non-active team members, and the unsupervised nature of these assignments makes it difficult for fair assessments \cite{marzanoClassroomManagementThat2003}. FlashHack effectively mitigated this issue due to close monitoring during the assessment; all team members were actively engaged, ensuring fairness in the evaluation.
    \item Feedback from students indicated that the majority found the Hackathon process and experience positive, with only a small number citing time limitations as a concern.
    \item Students did not need to commit a week to finish the homework. In most project assignments, students work independently and only receive instructor comments and support as needed. The instructor gave students prompt and specific questions during FlashHack. The students collaborated with teammates to complete the project within the given timeframe. 
    \item The instructors were pleased that the assessment of all teams was completed in a single day, which was not burdensome, as evaluating several projects typically requires more time for instructors.   %explain how this approach helps instructors in better assessment - quick assessment and avoid sleeping partners. 
\end{itemize}

\subsubsection{What could be Improved?}
\begin{itemize}
    \item As observed in the qualitative study, some students expressed concerns regarding time constraints, suggesting the possibility of extending the Hackathon duration to seven or eight hours, with intermittent breaks, to alleviate this issue.
    \item Implementing structured sections, such as specific questions for writing reflections, could aid in systematically compiling reflections post Hackathon.
    \item Collaboration with industry-connected institutes to procure relevant datasets and involve industry professionals in the evaluation process could enhance the authenticity and relevance of the Hackathon experience.
    \item Mindful team creation by instructors, considering factors such as gender and personality traits (e.g., introversion, extroversion) to ensure diverse and complementary skill sets within each team.
    \item Penalty for overusing Socratic questions or hints during the Hackathon will encourage independent problem-solving and prevent potential misuse.
    \item Assigning the dataset to teams before the Hackathon to allow ample time for data comprehension, hence preventing misinterpretation and the subsequent misuse of machine learning models.
\end{itemize}

%% file: limitations.tex
% !TeX root = main.tex

% generalizble 

% repeat this exercise again, correlate this with overall course performance 

% check retention in hackathon interest by gauging if they are enrolling in hackathons later outside academia

%% file: conclusion.tex
% !TeX root = main.tex
\section{Conclusion}
\label{sec:conclusion}

In conclusion, the \textit{FlashHack} Hackathon designed and executed by us as an assessment tool provided a dynamic and engaging platform for students to enhance their machine learning skills through hands-on experience. The thematic analysis of participant feedback revealed several key themes, including positive learning experiences, the development of problem-solving skills, the importance of teamwork, and the challenges faced. The approach also aided instructors in quick assessment as compared to project based learning assessments. By addressing the areas of improvement mentioned in discussion, such as extending the duration of the Hackathon, having industry relevant datasets and mentors, and with mindful team creation, the Hackathon can be further refined and replicated by others. FlashHack can be a valuable assessment tool for other educators adapting the suggested improvements. It combines practical application with real-world problem-solving, making the learning process both fun and engaging for students. This approach not only enhances students' technical abilities but also fosters important soft skills such as teamwork and time management. By replicating this model, educators can create an enriching educational experience that motivates and excites students while effectively assessing their knowledge and skills in a practical setting.